\documentclass[12pt,a4paper,headings=small,oneside]{scrartcl}

\usepackage{amsmath, amssymb, amsfonts}
\usepackage{amsthm}
\usepackage{mathtools}
\usepackage{graphics}
\usepackage{epstopdf}
\usepackage{natbib}
\usepackage{bbm}
\usepackage{bm}
\usepackage{setspace}
\usepackage{color}
\usepackage[hyphens]{url}
\usepackage{hyperref}

\newtheorem{theorem}{Theorem}

\newtheorem{proposition}[theorem]{Proposition}

\theoremstyle{definition}

\newtheorem{example}[theorem]{Example}

\newtheorem{remark}[theorem]{Remark}

\newcommand{\cd}{\stackrel{\mathcal{D}}{\longrightarrow}}

\newcommand{\Ex}{\mathbb{E}} 
\newcommand{\Var}{\operatorname{Var}} 
\newcommand{\Cov}{\operatorname{Cov}} 
\newcommand{\ind}{\mathbbm{1}} 

\begin{document}

\title{A Pareto tail plot without moment restrictions}
\author{Bernhard Klar\footnote{ bernhard.klar@kit.edu} \\
\small{Institute of Stochastics,} \\
\small{ Karlsruhe Institute of Technology (KIT), Germany.}
}

\date{\today }
\maketitle

\begin{abstract}
We propose a mean functional which exists for any probability distributions, and which characterizes the Pareto distribution within the set of distributions with finite left endpoint.
This is in sharp contrast to the mean excess plot which is not meaningful for distributions without existing mean, and which has a nonstandard behaviour if the mean is finite, but the second moment does not exist.
The construction of the plot is based on the so called {\em principle of a single huge jump}, which differentiates between distributions with moderately heavy and super heavy tails.
We present an estimator of the tail function based on $U$-statistics and study its large sample properties. The use of the new plot is illustrated by several loss datasets.
\end{abstract}

{\bfseries Keywords:} Pareto distribution, heavy tails, infinite mean, tail index,  single huge jump, regularly varying distribution.

\section{Introduction} \label{sec1}
Pareto distributions are probably the most important and widely used class of heavy-tailed distributions. A possible parameterization is
\begin{align}\label{pareto-dist}
	F(x;\alpha) &= 1-(x_m/x)^\alpha, \quad  x\geq x_m>0,
\end{align}
where $\alpha>0$. The generalized Pareto distribution (GPD) for $\xi> 0$ is parameterized by
\[
G(x;\beta,\xi)=1-(1+\xi x/\beta)^{-1/\xi}, \quad  x >0,
\]
where $\beta>0$. For $\xi>0$, a GPD can be converted to the form in (\ref{pareto-dist}) by a location-scale transformation.
For $\alpha\leq 1$ (i.e. $\xi\geq 1$), the mean is infinite; if  $1<\alpha\leq 2$ (i.e. $1/2\leq \xi<1$), the mean exists, but the variance is infinite.

A main tool for determining the adequacy of the (generalized) Pareto distribution as a model for the tails of sample data is the mean excess (ME) function, also known as the mean residual life function in reliability theory. It is given by
\[
M(u)= \mathbb{E}\left[X-u|X>u\right], \quad u>0,
\]
where $X$ is a positive random variable with $\Ex X<\infty$. 
In fact, the GPD class with $\xi<1$ is characterized by the linearity of the ME function \citep{EKT:1997}. 
The tail behavior of sample data by visual means can then be explored by the ME plot, i.e. a plot of the empirical coun\-ter\-part of the ME function.
Given an independent and identically distributed (iid) sample $X_1,\ldots,X_n\sim X$, the empirical ME function is defined by
\[
\hat{M}(u) = \frac{\sum_{i=1}^n (X_i-u) \ind\left(X_i>u\right)}{\sum_{i=1}^n \ind\left(X_i>u\right)},  \quad u>0.
\]
If $X_{(1)}\leq \ldots\leq X_{(n)}$ denote the order statistics of $X_1,\ldots,X_n$, one typically plots the pairs
$(X_{(k)}, \hat{M}(X_{(k)})$ for $1< k\leq n$, see \cite{GR:2010} or \cite{DG:2016}. 
An unfortunate feature of the ME plot is that it is well-defined only for distributions with finite expectation. Indeed, \cite{GR:2010} showed that the ME plot converges to a random curve in the case $\xi>1$, which also holds in the case $\xi=1$ after suitable rescaling. Thus, the ME plot is inconsistent when $\xi\geq 1$. Consequently, knowledge about the finiteness of the mean is required, a task that faces fundamental difficulties \citep{RO:2004}, although there are some attempts to construct tests for or against the existence of a finite mean \citep{FE:2013,TR:2016}. For statistical inference (e.g. confidence bounds), a finite second moment, i.e. $\xi<1/2$ for the generalized Pareto distribution, is required to obtain a normal limit (see \cite{DG:2013} for a thorough discussion).

There are several other plotting tools used in connection with heavy tails and extreme values, for example the Pareto QQ plot, plots of estimators of the extreme value index like the Hill plot, see, e.g., \cite{DR:2012} or \cite{DG:2013}.

The chasm between Pareto distributions with and without existing first moment (the latter are called extremely heavy-tailed) has been addressed in several papers in the last years. For example, \citet{CH:2023} recently showed that having exposures in multiple iid extremely heavy-tailed Pareto losses is worse than having just one Pareto loss of the same total exposure. More formally, they showed that a convex combination of such Pareto losses is larger in the usual stochastic order than a single infinite mean loss.
This penalizes diversification, which is impossible if the expectation is finite. 
A special case of this result has already appeared in \citet{EMS:2002}. For more results along these lines, see \citet{CH:2023}.

The importance of loss models without finite variance or even finite mean has been demonstrated by the thorough analysis of various datasets.
For an impressive list of such examples, see \citet{CH:2023}. As another example, \citet{CT:2020} concluded that the sizes of pandemics, properly transformed into a distribution with unbounded support, have infinite mean.
In view of these examples, it is extremely unsatisfactory that the ME plot breaks down in this important range of parameter values where variance or mean does not exist.
To circumvent these problems, we propose an alternative mean functional that exists for arbitrary probability distributions and characterizes the Pareto distribution within the set of distributions with finite left endpoint.
Specifically, assume that $X,X_1,X_2$ are iid random variables from an absolutely continuous distribution with support $[x_m,\infty)$. Then, the function 
\begin{align} \label{tail-funct}
	t_X(u) &= t(u) = \Ex \left[ \frac{|X_1-X_2|}{X_1+X_2} \Big | \min\{X_1,X_2\}\geq u \right]
\end{align}
is constant if and only if $X$ has the distribution function $F(\cdot;\alpha)$ for some $\alpha>0$.

\begin{figure}
	\centering
	\includegraphics[width=\textwidth]{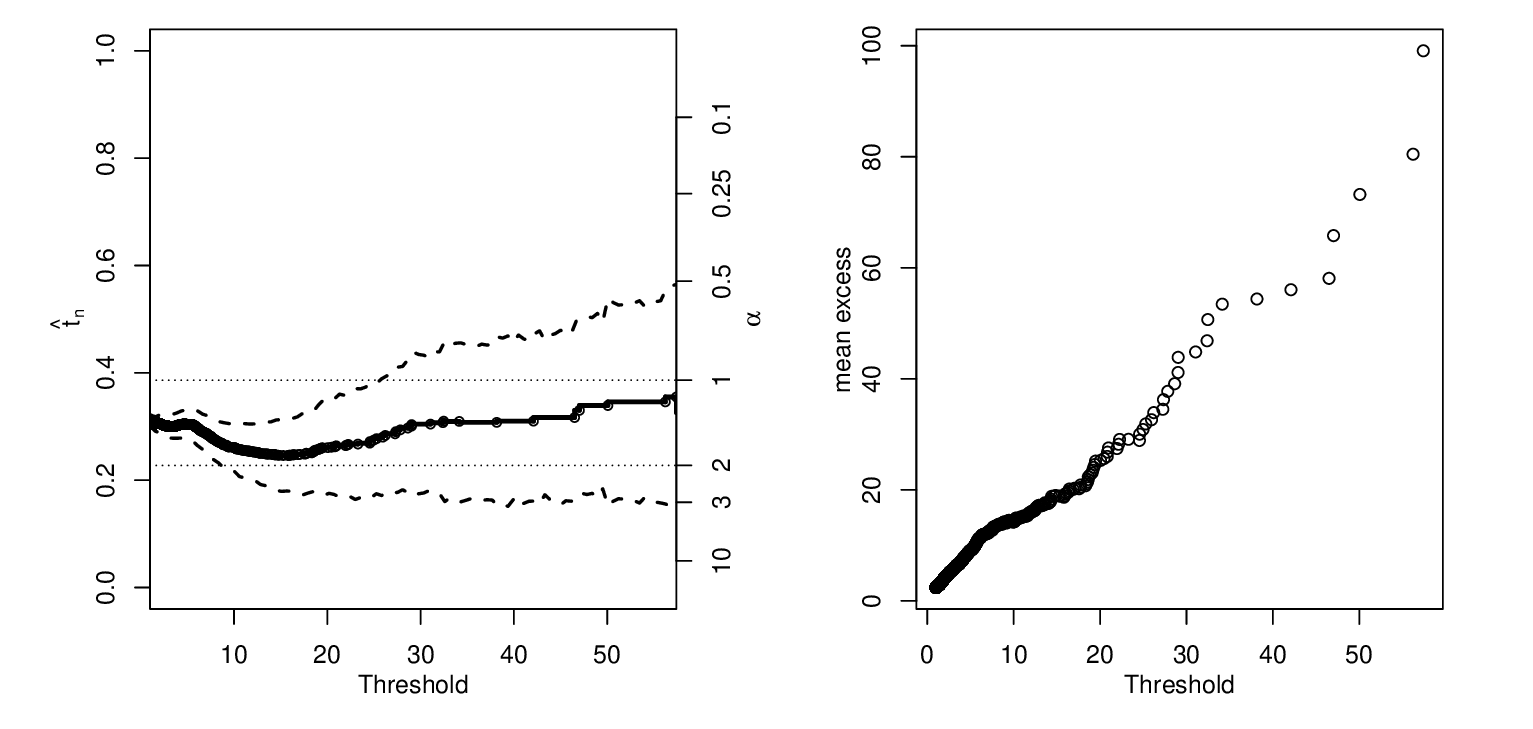}
	\caption{Plots for Danish fire insurance data.
		Left panel: Graph of the new Pareto tail function; the right scale indicates the corresponding shape parameter $\alpha$ under Pareto model.
		Right panel: Mean excess function.
		\label{fig3}}
\end{figure}

Let us illustrate the use of the plots with two examples.
First, we consider large fire insurance claims in Denmark, available as dataset \verb+danish+ in the R package \verb+evir+ \citep{PN:2018}.
The ME plot of this dataset with sample size $2167$ is shown in the right panel of Figure \ref{fig3}. The ME function shows an upward trend, indicating a heavy-tailed distribution. Since it follows a reasonably straight line, we may assume that the data follows a (generalized) Pareto distribution with positive shape parameter $\alpha$.
For this dataset, a detailed  analysis using methods from extreme value theory (EVT) is available \citep{MN:1997,RE:1997,MF:2015}, 
indicating a heavy-tailed distribution with a tail index $\alpha$ less than 2. Thus, the usual asymptotic results for the ME function assuming a finite second moment are not valid, and its interpretation is difficult.
The left panel of Figure \ref{fig3} shows our new Pareto tail plot. The graph is more or less horizontal with values between $0.25$ and $0.35$ (left axis), corresponding to values of $\alpha$ between $1.8$ and $1.1$ (right axis). Notice the bulge for thresholds $u$ between 5 and 15. For the specific values $u=5,10,15$, we get values of $0.30, 0.26, 0.25$, corresponding to $\alpha=1.40, 1.70, 1.82,$ respectively.
Looking at the width of the (pointwise) confidence intervals given by the dashed lines, we see that the data are compatible with Pareto models at each of these values.
These results are in good agreement with the cited literature:
Table 1 in \cite{MN:1997} gives estimates for $\alpha$ between $1.4$ and $2.0$; \cite{RE:1997} concludes ``based on an amalgam of the QQ, Hill and moment plots, we settle on an estimate of $\alpha=1.4$''.

\begin{figure}
	\centering
	\includegraphics[width=\textwidth]{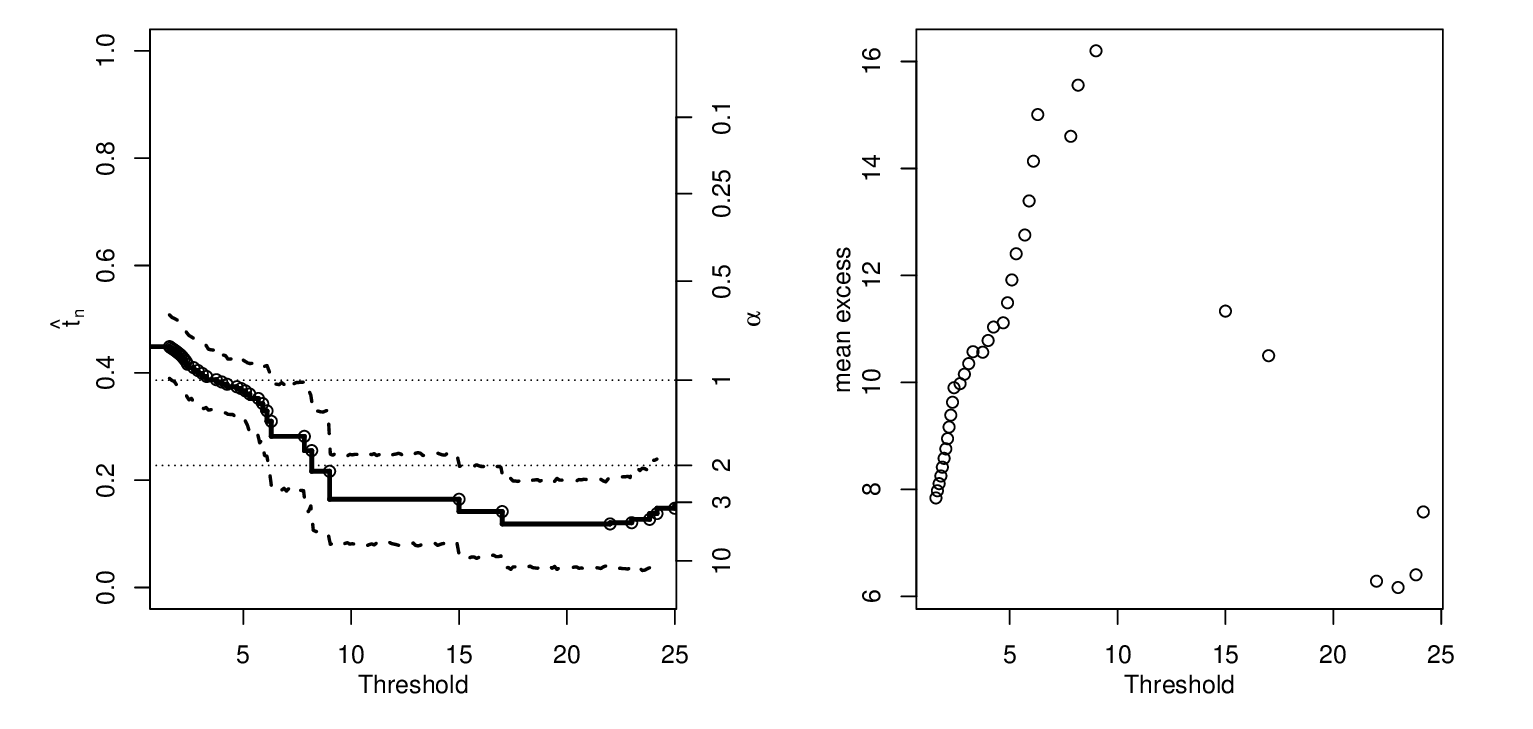}
	\caption{Plots for wind catastrophe losses.
		Left panel: Graph of of the new Pareto tail function; the right scale indicates the corresponding shape parameter $\alpha$ under Pareto model.
		Right panel: Mean excess function (not meaningful for $\alpha\leq 1$).
		\label{fig4}}
\end{figure}

Our second example considers a classical dataset of wind catastrophes taken  from \citet[p. 64]{HK:1984}. It represents 40 losses (in million U.S. dollars) due to wind-related disasters. Data are reported to the nearest million, including only losses of 2 million or more.
\cite{BS:2003} and \cite{RI:2009} proposed goodness-of-fit tests for the Pareto model and applied them to the de-grouped wind catastrophes data, and concluded that there were no evidence against the model.
In addition to the tests used in these articles, there are a variety of formal tests for Pareto models, see \cite{CD:2019} for an overview.
Estimates for $\alpha$ under this model range from $0.605$ to $0.791$.
The right panel of Figure \ref{fig4} again shows the mean excess function, which, however, is meaningless for values of $\alpha$ less than 1.
The left panel shows the new Pareto tail plot. The graph decreases, starting at $0.45$ for the full dataset and decreasing to $0.33$ for $u=6$ and $0.16$ for $u=10$, using only the 15 and 10 largest claims, respectively. Under a Pareto model, these values would correspond to $\alpha=0.79, 1.26$ and $2.89$, but the plot clearly argues against the validity of this model.
This is not a contradiction to the above remark: the fact that the goodness-of-fit tests do not reject the hypothesis of a Pareto distribution does not prove that the hypothesis holds. 

The article is organized as follows. Section 2 formally presents the new characterization of the Pareto distribution and states some fundamental properties. Section 3 introduces and discusses the intimately connected {\em principle of a single huge jump}, which differentiates between distributions with moderately heavy and super heavy tails.
In Section 4, an estimator of the tail function based on $U$-statistics is introduced, and its large sample properties are analyzed. Section 5 analyzes three additional datasets. Section 6 concludes the article.

\section{A new characterization of the Pareto distribution} \label{sec-pareto-plot}

In an abstract, \cite{SR:1965} indicated a characterization of the Pareto distribution with distribution function $F(x,\alpha)=1-(x_m/x)^\alpha$ for $x\geq x_m>0$, where $\alpha>0$, as follows. Let $F$ be an absolutely continuous distribution function with $F(x_m)=0$, and $X~\sim F$. Let $X_{(1)}\leq \ldots\leq X_{(n)}$ denote the order statistic of a sample of size $n$ from $F$. Then, $X_{(1)}$ and $(X_1+\ldots+X_n)/X_{(1)}$ are independent if and only if $X$ follows the Pareto distribution. This leads to the following result.

\begin{theorem} \label{pareto-char}
	Assume that $X,X_1,X_2\sim F$ are iid random variables, where the distribution function $F$ is absolutely continuous with $F(x_m)=0$. Then, the function $t_X:[x_m,\infty)\to [0,1]$ defined in (\ref{tail-funct}) satisfies
	\begin{align*}
		t_{X}(u) &= t_X(x_m), \quad \text{for all } u\geq x_m,
	\end{align*}
	if and only if $X$ is Pareto distributed, i.e. $X\sim F(\cdot,\alpha)$ for some $\alpha>0$.
	
	If $X\sim F(\cdot,\alpha)$, then, independent of $x_m$,
	\begin{align}\label{t-pareto}
		\widetilde{t}_{\alpha} &= t_X(x_m) = \left\{
		\begin{array}{cc}
			2\alpha \sum_{k=1}^{\alpha-1} \frac{(-1)^{\alpha+1-k}}{k} + (-1)^{\alpha+1}\, 2\alpha\log 2 - 1, & \alpha\in\mathbb{N}, \\
			\alpha \, \left( \Psi\left(\frac{\alpha+1}{2}\right) -\Psi\left(\frac{\alpha}{2}\right) \right) - 1,
			& \alpha>0, \alpha\notin\mathbb{N},
		\end{array} \right.
	\end{align}
	where $\Psi(z)=\frac{d}{dz} \log \Gamma(z)$ denotes the digamma function.
\end{theorem}

For Pareto distributed random variables with shape parameter $\alpha>0$, the proof of Theorem \ref{pareto-char} in Appendix \ref{appendix-A} shows that
$\widetilde{t}_{\alpha}= 2\int_0^1 \alpha y^{\alpha-1}/(1+y)dy-1$. Using integration by parts, we obtain
\begin{align*}
	\widetilde{t}_{\alpha} &= 2\int_0^1 \frac{y^{\alpha}}{(1+y)^2} \,dy.
\end{align*}
This representation reveals that $\widetilde{t}_{\alpha}$ is strictly decreasing as a function of $\alpha$, with $\lim \widetilde{t}_{\alpha}=1$ for $\alpha\to 0$, and
$\lim \widetilde{t}_{\alpha}=0$ for $\alpha\to \infty$.
Specific values are given by $\widetilde{t}_{1/2}=\pi/2-1\approx 0.571, \widetilde{t}_{1}=2\log 2-1\approx 0.368$, $\widetilde{t}_{2}=3-4\log 2 \approx 0.227$ and $\widetilde{t}_{3}=-4+6\log 2 \approx 0.159$.

Since the conditional distribution of a Pareto-distributed random variable $X$, given $X>y$ (where $y>x_m$), is a Pareto distribution with the same $\alpha$ but with minimum  $y$ instead of $x_m$, we have $t_{X|X>y}(u)=t_X(u)$ for $u\geq y$.
Furthermore, the function $t$ satisfies $t_{cX}(u)=t_X(u/c)$ for $c>0$ and $u\geq cx_m$. It has the following interpretation (cp. \cite{AL:2017}): if both $X_1$ and $X_2$ contribute equally to the sum $X_1+X_2$, then $t$ should eventually obtain values close to 0; if only one of the variables tends to be of the same magnitude as the whole sum, then $t$ is close to 1 for large $u$.
However, in contrast to the functional $g$ considered in \cite{AL:2017} and \cite{IK:2023}, this evaluation is performed only if both $X_1$ and $X_2$ are large, and not already if the sum is large. This interpretation will be explored further in the next section.

\section{The principle of a single huge jump} \label{sec-huge-jump}

Let $X_1$ and $X_2$ be independent random variables, having the same Weibull distribution with shape parameter $k$.
To illustrate the \emph{principle of a single big jump}, \cite{FK:2013} considered the distribution of the random
variable $X_1/X_1+X_2$ conditional on the sum $X_1+X_2=u$ for increasing values of $u$.
They showed that for $k<1$, i.e. for heavy-tailed distributions, this distribution converges for $u\to \infty$ to $(\delta_{0}+\delta_{1})/2$, where $\delta_x$ denotes the Dirac measure in $x$. 
Here, the distribution (function) $F$ is said to be {\em heavy-tailed} if
\[
\int_{-\infty}^{\infty} e^{\lambda x} F(dx)=\infty \quad \text{for all } {\lambda>0},
\]
otherwise F is said to be light-tailed \citep{FK:2013}.
For Weibull distributions with $k>1$, i.e. light-tailed distributions, and as $u\to \infty$, the distribution converges to $\delta_{1/2}$.
For $k=1$, i.e. for the exponential distribution, there is no concentration of mass. 
Further results in this direction have been obtained by \cite{LE:2015}. 

Instead of $X_1/u$, we can consider the distribution of the random variable $(X_{(2)}-X_{(1)})/u$ conditional on the sum $X_1+X_2=u$, which converges to $\delta_1$ for $k<1$, and to $\delta_0$ for $k>1$ as $u\to \infty$. The first behavior is typical for heavy-tailed variables, and is an example of the principle of a single big jump: if the sum $X_1+X_2$ is large, then one of the variables is large compared to the other. For light-tailed distributions, however, both of the variables $X_1$ and $X_2$ contribute equally, and the difference is small compared to the sum. 

Now let $X_1$ and $X_2$ be non-negative and iid random variables. Assume that $X_1$ has unbounded support $(x_m,\infty)$ and density function $f$. Let us consider a similar setting as above, but where we condition on $\min\{X_1,X_2\}=u$ instead of $X_1+X_2=u$.
We will see that this change leads to a formally similar partition within the distributions with heavy tails.
Since the joint density of the order statistics
$(X_{(1)},X_{(2)})$ is given by $f_{()}(x_1,x_2)=2f(x_1)f(x_2)$ for $x_1<x_2$, the joint density of
$(Y_1,Y_2)=(X_{(1)},(X_{(2)}-X_{(1)})/(X_{(1)}+X_{(2)}))$ is
\begin{align*}
	g(y_1,y_2) &= 2f(y_1) \, f\left(\frac{y_1(1+y_2)}{1-y_2}\right) \, \frac{2y_1}{(1-y_2)^2}, \quad y_1>x_m, \, 0<y_2<1.
\end{align*}
Division by the density of the first order statistics yields the density of the random variable $Z=(X_{(2)}-X_{(1)})/(X_{(1)}+X_{(2)})$ conditional on the minimum $X_{(1)}=u$ as
\begin{align} \label{cond-dens}
	g(z|u) &= \frac{2u \, f\left(\frac{u(1+z)}{1-z}\right)} {(1-z)^2 \, \bar{F}(u)}, \quad 0<z<1, \, u>x_m.
\end{align}

\begin{figure}
	\centering
	\includegraphics[width=\textwidth]{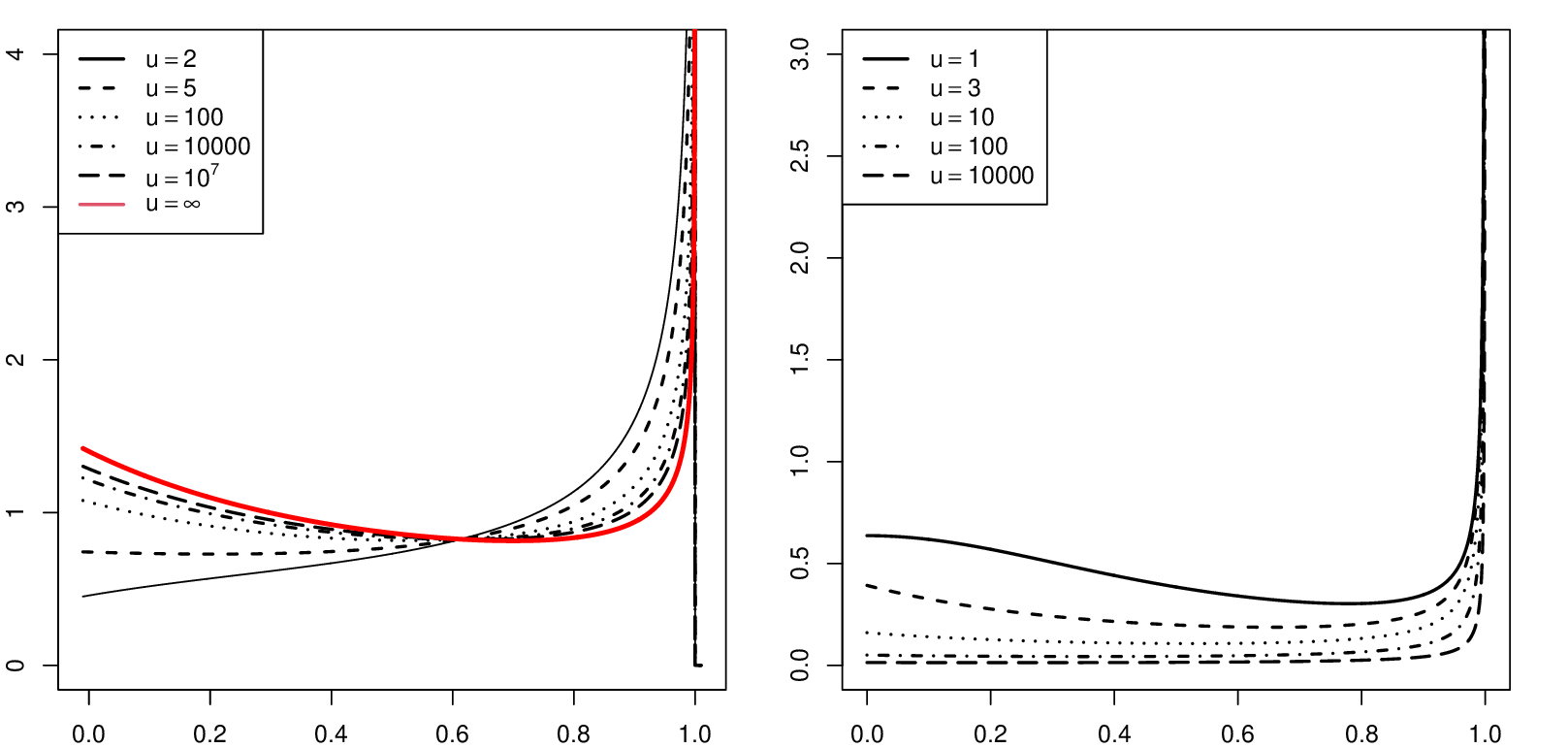}
	\caption{Plots of the conditional density $g(z|u)$ in (\ref{cond-dens}) for log-gamma (with $\alpha=0.7,\beta=2$, left panel) and log-Cauchy distribution (right panel) for increasing values of the boundary. 
		Left panel: $g(z|u)$ (in black) converges to $\tilde{g}_{\alpha}(z)$ (in red) as $u\to\infty$. \\
		Right panel: $g(z|u)$ converges to $\delta_1$ as $u\to\infty$.
		\label{fig1} } 
\end{figure}

\begin{example} \label{ex-huge-jump}
	\begin{enumerate}
		\item[a)]
		For the Weibull distribution with shape parameter $k$, we obtain
		\begin{align*}
			g(z|u) &= 2ku^k (1+z)^{k-1} (1-z)^{-(k+1)}
			\exp\left\{-u^k \left[ \left((1+z)/(1-z)\right)^k - 1 \right] \right\}
		\end{align*}
		for $0<z<1, u>0$. It follows that the conditional distribution converges to a one-point distribution in 0 as $u\to\infty$ for arbitrary $k$. Thus, for heavy-tailed Weibull distributions, the distribution of $Z$ conditional on $X_{(1)}=u$ behaves completely different from conditioning on $X_1+X_2=u$ as $u\to\infty$.
		\item[b)]
		For the Pareto distribution with shape parameter $\alpha$, the conditional density is
		\begin{align*} 
			g(z|u) &= \tilde{g}_{\alpha}(z) = 2\alpha \, (1+z)^{-(\alpha+1)} (1-z)^{\alpha-1}, \quad 0<z<1,
		\end{align*}
		for all $u>x_m$. The conditional distribution does not depend on $u$, in agreement with Theorem \ref{pareto-char}.
		Thus, there is no concentration of mass and the family of Pareto distributions takes over the role of the exponential distributions in the first scenario.
		\item[c)]
		Denote by $\cal{R}_{\alpha}$ the class of regularly varying distributions, where $\bar{F}(x)=L(x)/x^\alpha$ with $\alpha>0$ and $L(\cdot)$ is slowly varying, i.e. $\lim_{x\to\infty} L(tx)/L(x) = 1$ for $t>0$. Assume that $F$ has an ultimately decreasing density $f$, i.e. there exists $c$ such that $f$ is decreasing on $[c,\infty)$. This ensures that $f(x) = \alpha \tilde{L}(x) / x^{\alpha+1}$ with $\tilde{L}(x) \sim L(x)$ as $x\to\infty$ \citep[Th. A.3.7]{EKT:1997}. Here, $f(x)\sim g(x)$ as $x\to\infty$ means that $\lim_{x\to \infty} f(x)/g(x)=1$. Then, the conditional density is  
		\begin{align*}
			g(z|u) &=   2\alpha \, \frac{(1-z)^{\alpha-1}}{(1+z)^{\alpha+1}} \, \frac{\tilde{L}\left(\frac{u(1+z)}{1-z} \right)}{L(u)},  \quad 0<z<1,
		\end{align*}
		for all $u>x_m$. Therefore, $g(z|u)\sim \tilde{g}_{\alpha}(z)$ as $u\to \infty$. Thus, the limiting behavior of $g(z|u)$ for distributions in $\cal{R}_{\alpha}$ coincides with that of a Pareto distribution with the same shape parameter.
		
		The left panel in Figure \ref{fig1} shows graphs of $g(z|u)$ for the log-gamma distribution with density
		$$
		f(x) = \frac{\alpha^\beta}{\Gamma(\beta)}(\log x)^{\beta-1} x^{-\alpha-1}, \quad x\geq 1 \quad (\alpha,\beta>0),
		$$
		for increasing values of $u$. The conditional density  $g(z|u)$ of this regularly varying distribution converges to $\tilde{g}_{\alpha}(z)$ (plotted in red) as $u\to \infty$.
		
		\item[d)]
		The right panel in Figure \ref{fig1} shows graphs of $g(z|u)$ for the (standard) log-Cauchy distribution, a distribution with heavier tails than Pareto distributions. Here, the conditional distribution of $Z$ converges to a one point distribution in 1 as $u\to\infty$.
	\end{enumerate}
\end{example}

The last example of the log-Cauchy distribution shows an effect which can be called the \emph{principle of a single huge jump}: if each of the variables $X_1$ and $X_2$ takes values over a large boundary, then one of the variables will still typically be much larger than the other. Distributions that follow this principle, and hence $g(z|u) \to \delta_1$ as $u\to\infty$, could be called \emph{super heavy-tailed}.
There are other definitions of super heavy-tailed distributions in the literature \citep{AL:2009,CR:2009}; their connections to the definition given here require further investigation.
On the contrary, if we consider observations $X_1$ and $X_2$ over a large boundary for Weibull distributions,
then $X_1$ and $X_2$ contribute equally, and the difference is small compared to the sum. Thus, such distributions, where $g(z|u) \to \delta_0$ as $u\to\infty$, are (at most) moderately heavy-tailed. In between are the Pareto distributions for arbitrary shape parameters, where a limiting density of $g(z|u)$ exists without any concentration of probability mass.

The shape of the conditional density in Example \ref{ex-huge-jump} c) suggests that the limiting behavior of $t(u)$ as $u\to\infty$  for the class of regularly varying distributions is the same as for Pareto distributions. 
If $L$ is slowly varying, then $L(ux)/L(x)\to 1$ as $x\to\infty$ uniformly over compact $u$ sets \citep[Th. A.3.2]{EKT:1997}. 
Strengthening this property, we can indeed find the limiting behavior of $t$ for a subset of regularly varying distributions; the proof is given in Appendix \ref{appendix-B}.

\begin{theorem} \label{th-reg-var}
	Assume that $X,X_1,X_2\in \cal{R}_{\alpha}$ are iid random variables, where the distribution is absolutely continuous with an  ultimately decreasing density. 
	Assume further that $\lim_{x\to\infty}L(ux)/L(x)=1$ uniformly in $u$ on $(x_m,\infty)$. 
	Then, 
	\begin{align*}
		\lim_{u\to\infty} t_{X}(u) &= \widetilde{t}_{\alpha},
	\end{align*}
	where $\widetilde{t}_{\alpha}$ is defined in Theorem \ref{pareto-char}.
\end{theorem}

\section{An estimator of \texorpdfstring{\boldmath $t(u)$}{t(u)} based on \texorpdfstring{\boldmath $U$}{U}-statistics}

To estimate $t$ based on the order statistics $X_{(1)}\leq \ldots\leq X_{(n)}$ , define $U$-statistics
\begin{align*}
	U_n^{(1)}(u) &= \frac{2}{n(n-1)} \sum_{1\leq i<j\leq n} \frac{X_{(j)}-X_{(i)}}{X_{(i)}+X_{(j)}}  \, \ind\left(X_{(i)}\geq u\right),  \\
	U_n^{(2)}(u)
	&= \frac{2}{n(n-1)} \sum_{1\leq i<j\leq n}\ind\left( X_{(i)}\geq u\right)
	= \frac{2}{n(n-1)} \sum_{i=1}^n (n-i)\,\ind\left( X_{(i)}\geq u\right),
\end{align*}
where $\ind(B)$ is the indicator function of the event B.
A nonparametric estimator of $t(u)$ is then given by the ratio of these statistics:
\begin{align*}
	\hat{t}_n(u) &= \frac{U_n^{(1)}(u)}{U_n^{(2)}(u)}, \quad u\geq x_m.
\end{align*}
Note that it suffices to evaluate  $\hat{t}_n$ at the sample points, yielding
\begin{align*}
	\hat{t}_n\left(X_{(k)}\right) &= \frac{1}{\binom{n-k+1}{2}} \
	\sum_{k\leq i<j} \frac{X_{(j)}-X_{(i)}}{X_{(i)}+X_{(j)}}, \quad k=1,\ldots,n-1.
\end{align*}
Hence, $\hat{t}_n$ itself can be seen as a U-statistic, applied to the $n-k+1$ largest observations. For computational purposes, note that
\begin{align*}
	\hat{t}_n\left(X_{(k)}\right) &=
	\frac{n-k+2}{n-k} \, \hat{t}_n\left(X_{(k-1)}\right)
	- \frac{1}{\binom{n-k+1}{2}} \
	\sum_{j=k}^n \frac{X_{(j)}-X_{(k-1)}}{X_{(k-1)}+X_{(j)}}, \quad k=2,\ldots,n-1.
\end{align*}
This can be evaluated very quickly, starting with
$\hat{t}_n(X_{(n-1)})=(X_{(n)}-X_{(n-1)})/(X_{(n)}+X_{(n-1)})$.

\begin{figure}
	\centering
	\includegraphics[width=\textwidth]{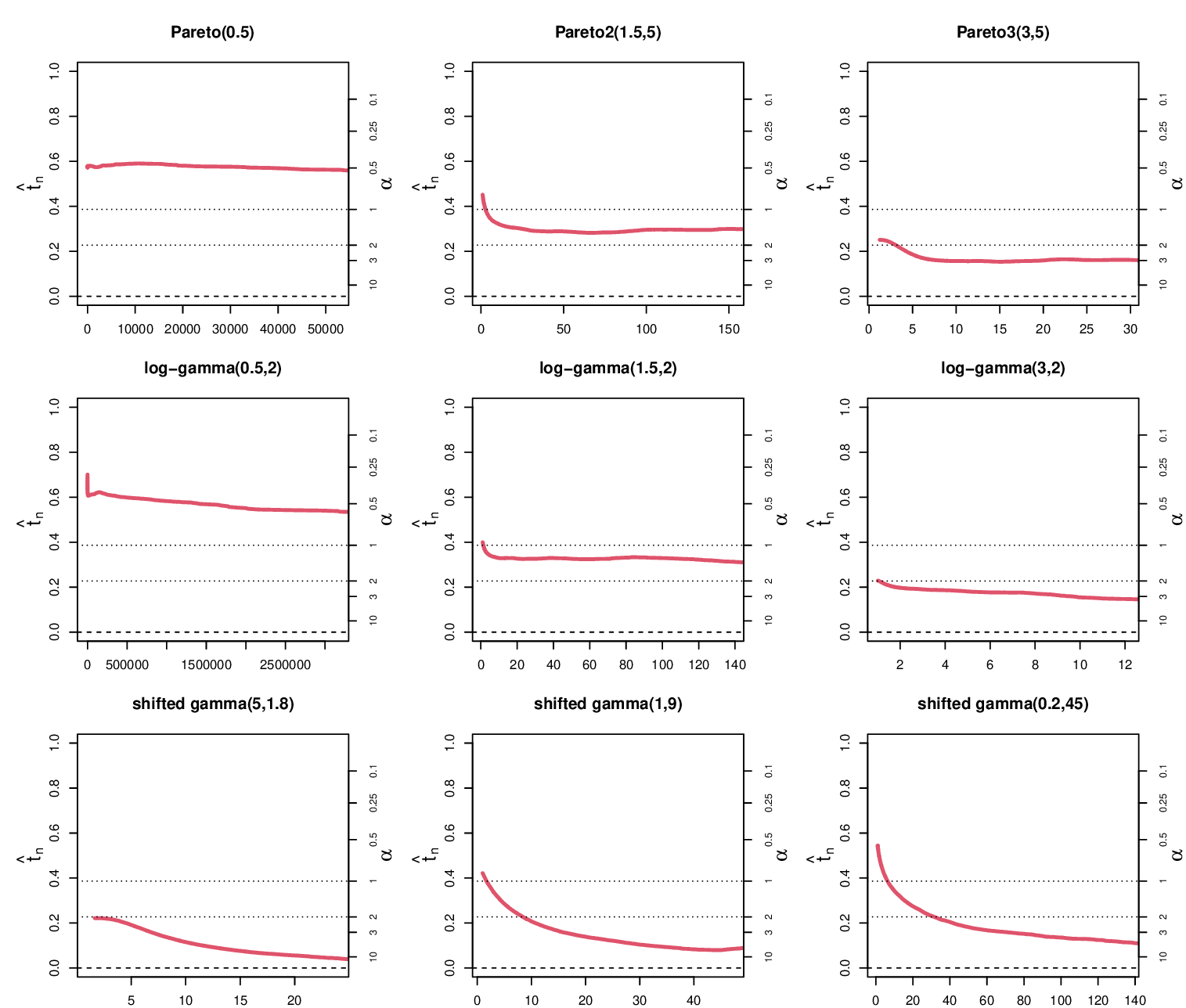}
	\caption{Graphs of $\hat{t}_n(u)$ for simulated samples from different distributions on $(1,\infty)$ with sample size $10~000$.
		Upper row: Pareto distributions of type 1,2,3 with shape parameter $0.5, 1.5, 3$.
		Middle row: Log-gamma distributions with $\alpha=0.5, 1.5, 3$ and $\beta=2$.
		Lower row: Shifted gamma distributions, each with expected value 10.
		\label{fig2}}
\end{figure}

\medskip
Figure \ref{fig2} shows the graphs of $\hat{t}_n(u)$ for simulated samples with sample size $10~000$ from different distributions on $(1,\infty)$. The bound $u$ is in the range from 1 to the 0.995-quantile of the sample. 
The two dotted horizontal lines in the plots are given by
$\widetilde{t}_1\approx 0.39$ and $\widetilde{t}_2\approx 0.23$ in (\ref{t-pareto}).
The top row shows the Pareto distribution with shape parameter $0.5$ and the Pareto distributions of type 2 and 3 with scale parameter  $\theta=5$ and shape parameter $\alpha=1.5$ and 3, respectively. Their densities are given by 
$$
f_2(x) = \frac{\alpha}{ \theta(1+(x-1)/\theta)^{\alpha+1} }, \quad
f_3(x) = \frac{\alpha((x-1)/\theta)^{\alpha-1}}{\theta \big(1+ \big((x-1)/\theta\big)^\alpha \big)^2}, \quad x>1,
$$
and satisfy the assumptions of Theorem \ref{th-reg-var}. As expected, $\hat{t}_n(u)$ is nearly constant after a short settling period.
The middle row shows log-gamma distributions with $\alpha=0.5,1.5,3$ and $\beta=2$. These regularly varying distributions don't satisfy the uniform convergence assumption in Theorem \ref{th-reg-var}. Nevertheless, $\hat{t}_n$ seems to converge to the limits corresponding to the shape parameter $\alpha$.
Finally, the bottom row shows shifted gamma distributions, whose conditional densities $g(z|u))$ converge to $\delta_0$, as it is the case for Weibull distributions. Inspection of the proof of Theorem \ref{th-reg-var} suggests that in this case, $t(u)$ converges to 0 as $u\to\infty$, and the plots support this suggestion.

\begin{remark}
	By the results in section \ref{sec-pareto-plot}, $\widetilde{t}_{\alpha}$ is strictly decreasing in $\alpha$. Thus, one can estimate the parameter $\alpha$ of the Pareto distribution using $\hat{t}_n$. More generally, one can estimate the tail index of a distribution with Pareto tails (for various possible definitions see \cite{FE:2020}, Assumptions A1-A8), based on the $k$ largest observations of a sample. This adds another method to the more than one hundred Pareto tail index estimators listed in \cite{FE:2020}.
\end{remark}

\subsection{Large sample properties of \texorpdfstring{\boldmath $\hat{t}_n$}{tn}} \label{sec5}

The $U$-statistic $U_{n}^{(l)}(u)$ are unbiased estimators of $\theta^{(l)}_u=\Ex[h^{(l)}(X_1,X_2;u)], l=1,2$, where the kernels $h^{(l)}$ of degree 2 are defined by
\begin{align*}
	h^{(1)}(x_1,x_2;u) = \frac{|x_1-x_2|}{x_1+x_2} \, \ind\left(\min\{x_1,x_2\}>u\right),  &\quad
	h^{(2)}(x_1,x_2;u) = \ind\left(\min\{x_1,x_2\}>u\right).
\end{align*}
By the strong law of large numbers for $U$-statistics \citep[p.~122]{LE:1990}, $U_{n}^{(l)}(u), l=1,2,$ and hence $\hat{t}_n(u)$ are strongly consistent estimators for $\theta^{(l)}_u$ and $t(u)$, respectively.
Using the general theory of $U$-statistics (see  \citet[p.~76]{LE:1990} or \citet[p.~132]{KB:1994})), we can derive the joint asymptotic distribution of $(U_{n}^{(1)}(u),U_{n}^{(2)}(u))$.

\begin{proposition} \label{prop4}
	For $l=1,2$, let
	\begin{align*}
		\psi^{(l)}(x_1,x_2;u) &= h^{(l)}(x_1,x_2;u)-\theta^{(l)}_u, \quad
		\psi^{(l)}_1(x_1;u) = \Ex\big[\psi^{(l)}(x_1,X_2;u)\big].
	\end{align*}
	Further, define
	\begin{align*}
		\eta^{(l)}_1(u)=\Ex\left[\left( \psi^{(l)}_1(X_1;u) \right)^2\right] \ (l=1,2), \quad
		\eta^{(1,2)}_1(u)=\Ex\left[\psi^{(1)}_1(X_1;u)\, \psi^{(2)}_1(X_1;u)\right].
	\end{align*}
	If $\eta^{(l)}_1(u)>0$ for $l=1,2$, then,
	\begin{align*}
		\sqrt{n} \left(
		\begin{pmatrix} U_{n}^{(1)}(d) \\ U_{n}^{(2)}(u) \end{pmatrix} -
		\begin{pmatrix} \theta^{(1)}_u \\ \theta^{(2)}_u \end{pmatrix} \right)
		& \cd N_2(\mathbf{0},4\Sigma_u),
		\quad \text{where } \Sigma_u =
		\begin{pmatrix}
			\eta^{(1)}_1(u) & \eta^{(1,2)}_1(u) \\
			\eta^{(1,2)}_1(u) & \eta^{(2)}_1(u)
		\end{pmatrix}.
	\end{align*}
\end{proposition}

Using Prop. \ref{prop4} and the delta method, we can derive the asymptotic behaviour of $\hat{t}_n(u)$.

\begin{theorem} \label{theorem4}
	Let $\nu_u=\theta^{(2)}_u>0$, and $\eta^{(l)}_1(u)>0$ for $l=1,2$. Then,
	\begin{align} \label{asymp-g-hat}
		\sqrt{n\nu_u} \left( \hat{t}_n(u) - t(u) \right)
		& \cd N\left(0, \sigma_u^2 \right),
	\end{align}
	where
	\begin{align} \label{sigma-d}
		\sigma_u^2 = \frac{4}{\nu_u} \left( \eta^{(1)}_1(u)
		- 2t(u) \, \eta^{(1,2)}_1(u) + t^2(u) \, \eta^{(2)}_1(u) \right).
	\end{align}
\end{theorem}

Next, we want to construct confidence intervals for $t(u)$.
Based on Theorem \ref{theorem4}, a confidence interval with asymptotic coverage probability $1-\gamma$ is given by
\begin{align*}
	\left[ \max\left\{ \hat{t}_n(u) - \frac{z_{1-\gamma/2}\, \hat\sigma_u}{ (nU_n^{(2)}(u))^{1/2}}, 0\right\},  \;
	\min\left\{ \hat{t}_n(u) + \frac{z_{1-\gamma/2}\, \hat\sigma_u}{ (nU_n^{(2)}(u))^{1/2}}, 1\right\} \right],
\end{align*}
where $z_{p}=\Phi^{-1}(p)$, and $\hat\sigma_u^2$ is an estimator of  the variance of $\sqrt{n\nu_d}\,\hat{t}_n(u)$, corresponding to $\sigma_u^2$ in (\ref{sigma-d}). A possible approach is to derive  unbiased estimators $\sigma_{U^{(l)}}^2 (l=1,2)$ and $\sigma_{U^{(1,2)}}^2$ of $\Var(U_{n}^{(l)}(u))$ and  $\Cov(U_{n}^{(1)}(u),U_{n}^{(2)}(u))$, respectively.
This can be done efficiently with complexity $O(n^2)$, see Appendix \ref{appendix2}.
Then, an estimator of $\sigma_u^2$ is given by
\begin{align} \label{hat-sigma-d}
	\hat{\sigma}_u^2 = \frac{n}{U_{n}^{(2)}(d)} \left(
	\sigma_{U^{(1)}}^2 - 2\hat{t}_n(u) \, \sigma_{U^{(1,2)}}^2
	+ \hat{t}_n^2(u) \, \sigma_{U^{(2)}}^2 \right).
\end{align}
Further possible estimators are the bootstrap variance estimator $\hat{\sigma}_{B,u}^2$ and the jackknife estimator $\hat{\sigma}_{J,u}^2$. Details can be found in \citet{SS:1992} and \citet{IK:2023}.

\subsection{Empirical coverage probability of confidence intervals for  \texorpdfstring{\boldmath $t(u)$}{t(u)}}

In this subsection, we empirically study the coverage probabilities of confidence intervals for $t(u)$ based on the three variance estimators introduced in section \ref{sec5}.
As distributions, we choose Pareto distributions with $x_m=1$ and different shape parameters $\alpha$. In all simulations, we use effective sample sizes, defined as follows: for given values of $\alpha$ and $u$, the total sample size $n$ was chosen such that $n\nu_u= nP(\min\{X_1,X_2\}>u)=n_{\text{eff}}$.
All computations have been performed using the statistic program R \citep{R:2023}.

The results for confidence level $1-\gamma=0.95, n_{\text{eff}}=20, u=2$ and varying $\alpha$ are given in Table \ref{tab1}. For this small sample size, all intervals are anticonservative, i.e. they have coverage probability less than 0.95. The jackknife estimator $\hat{\sigma}_{J,u}^2$ performs slightly better than the competitors for $\alpha\geq 0.5$. For $\alpha=0.2$, the coverage probability with $\hat{\sigma}_u^2$ is as low as $0.90$, while the other two estimators work well.

\begin{table}
	\centering
	\setlength{\tabcolsep}{12pt}
	\begin{tabular}{rrrr}
		\hline
		$\alpha$ & $\hat{\sigma}_{u}^2$ & $\hat{\sigma}_{B,u}^2$ & $\hat{\sigma}_{J,u}^2$ \\ \hline
		0.2 & 90.0 & 93.2 & 93.2 \\
		0.5 & 91.6 & 92.3 & 93.4 \\
		1.0 & 92.0 & 92.1 & 93.3 \\
		2.0 & 93.0 & 93.0 & 93.7 \\
		3.0 & 93.5 & 93.4 & 93.8 \\    \hline
	\end{tabular}
	\setlength{\tabcolsep}{6pt}
	\caption{Empirical coverage probability of 0.95-confidence intervals for $t(u)$ based on different estimators for effective sample size $n_{\text{eff}}=20, u=3$ and varying $\alpha$.} \label{tab1}
\end{table}

Table \ref{tab2} shows the results for $1-\gamma=0.95, \alpha=1, u=2$ and increasing sample sizes. For $n_{\text{eff}}=40$, $\hat{\sigma}_{J,u}^2$ still has the edge over the competitors.
For $n_{\text{eff}}=80$ or larger, all intervals work very well.
Since computing time of $\hat{\sigma}_{J,u}^2$ and of $\hat{\sigma}_{B,u}^2$, which was computed with 999 bootstrap replications, becomes a problem for large samples, we recommend to use $\hat{\sigma}_{J,u}^2$ for small to moderate samples, and to switch to $\hat{\sigma}_{u}^2$ for large samples sizes.

\begin{table}
	\centering
	\setlength{\tabcolsep}{12pt}
	\begin{tabular}{rrrrr}
		\hline
		$n_{\text{eff}}$ & $\hat{\sigma}_{u}^2$ & $\hat{\sigma}_{B,u}^2$ & $\hat{\sigma}_{J,u}^2$ \\ \hline
		10.0 & 87.9 & 89.4 & 91.1 \\
		20.0 & 91.8 & 91.7 & 93.0 \\
		40.0 & 93.2 & 93.0 & 93.7 \\
		80.0 & 94.5 & 94.5 & 94.7 \\
		160.0 & 94.9 & 94.8 & 95.0 \\  \hline
	\end{tabular}
	\setlength{\tabcolsep}{6pt}
	\caption{Empirical coverage probability of 0.95-confidence intervals for $t(u)$ based on different estimators for $\alpha=1, u=2$ and increasing effective sample size.} \label{tab2}
\end{table}

\section{Empirical illustrations}

In this section, we describe and analyze three additional datasets:
{\em French marine losses (2006), wildfire suppression costs (1995), costs of nuclear power accidents (2016)}.
The first two datasets were also considered in \cite{CH:2023}.

\subsection{French marine losses} \label{ss-marine}

\begin{figure}
	\centering
	\includegraphics[width=\textwidth]{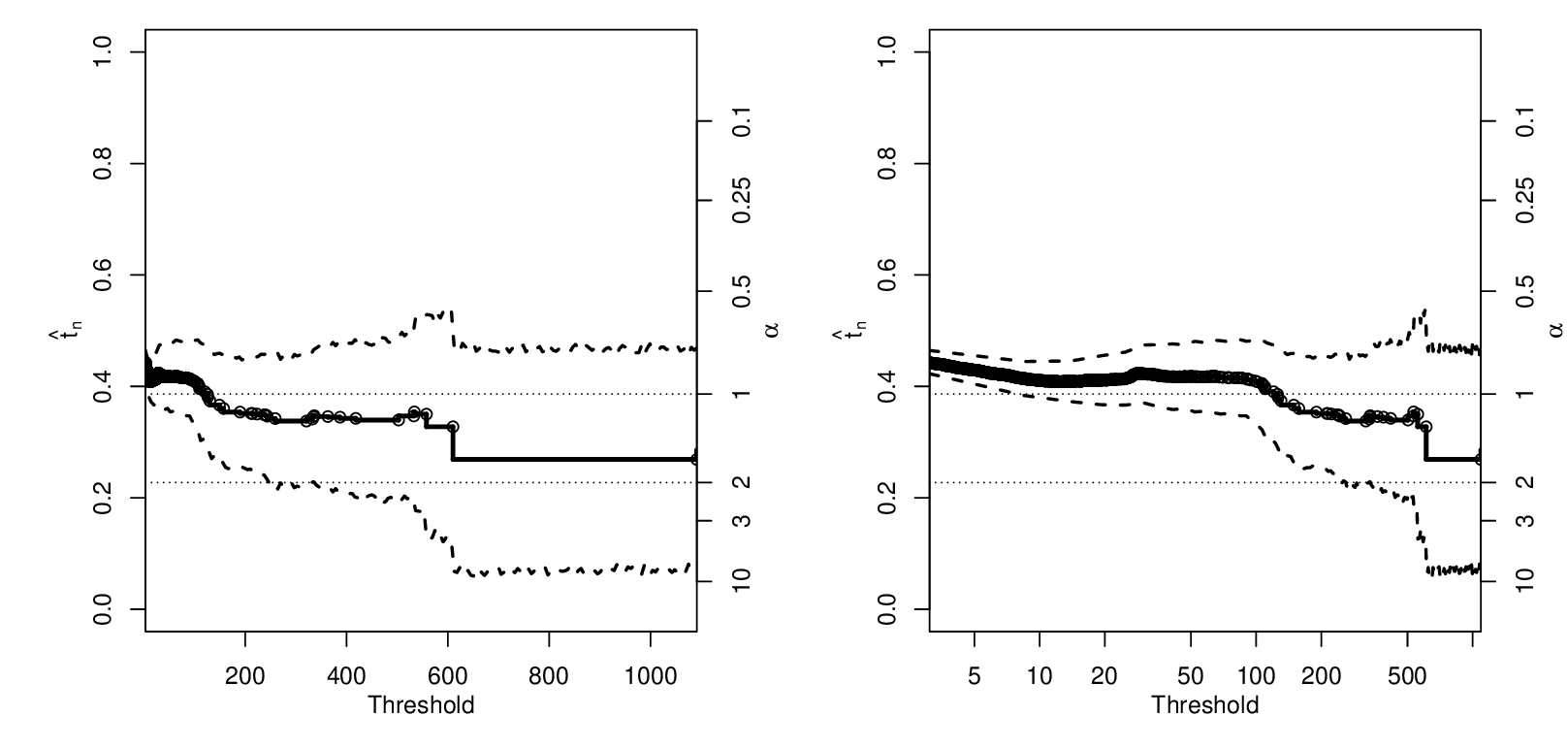}
	\caption{Plots for dataset on French marine losses, $n=657$. \\
		Left panel: Graph of $\hat{t}_n(u)$; the right scale indicates the corresponding shape parameter $\alpha$ under the assumption that the data follows a Pareto distribution.  \\
		Right panel: horizontal axis in log scale.
		\label{fig6}}
\end{figure}

The marine losses dataset from the insurance data repository \verb+CASdatasets+, available at \verb+http://cas.uqam.ca/+, comes from a French private insurer and includes 1274 marine losses between January 2003 and June 2006. We consider only the paid amount of the claims (which have been rescaled to mask the actual losses) over 3 monetary units. This results in a sample of size $657$. The Pareto tail plot on the original scale is shown in the left panel of Figure \ref{fig6}; in the right panel, the horizontal axis is log-scaled for better visualization.
For smaller thresholds up to 100, the function $\hat{t}_n$ takes values slightly above 0.4, corresponding to a shape parameter below 1.
For example, for $u=20,50,100$, which corresponds to using the $167, 72, 37$ largest data points, the values of $\hat{t}_n(u)$ are given by $0.411, 0.418,0.408$, in turn corresponding to values of $\alpha$ around $0.91, 0.89, 0.92,$ respectively. Then, $\hat{t}_n$ decreases to values around 0.35. For example, $\hat{t}_n(300)=0.338$, which corresponds to $\alpha=1.21$.
However, in this region, the confidence intervals indicate a high uncertainty in the estimates, not contradicting the assumption of a Pareto model with a shape parameter below 1.
While plotting the mean excess function makes no sense for this dataset, \cite{CH:2023} present a Hill plot of this dataset. Using a threshold around the top 5\% of the order statistics of the data yields a Hill estimate of $0.916$ for the parameter $\alpha$, which is in good agreement with our analysis.

\subsection{Wildfire suppression costs} \label{ss-wildfire}

The wildfire dataset contains $10\,915$ suppression costs (in Canadian dollars) collected in Alberta from 1983 to 1995.
We only consider costs above 1000 dollars and divide them by 1000, resulting in a dataset of size 6599, with minimum value of 1. Therefore, the Pareto tail plot in the left panel  in Figure \ref{fig5} shows the costs in 1000 dollars on the x-axis, while it is log-scaled in the right panel.
Specific values are given in Table \ref{tab3}. For small and medium thresholds, the shape parameter is estimated to be less than 1.
For $u=25$, which is approximately the upper $5\%$ quantile of the complete dataset, the estimate of $\alpha$ is 0.85, essentially identical to the value given in \cite{CH:2023}. This suggests a distribution with infinite mean. 
However, for larger thresholds, $u=150$ and above, the function $\hat{t}_n$ drops well below $0.368$ and becomes stable, with values around 0.34, corresponding to $\alpha=1.20$. For $u=150$, the estimate is still based on 142 observations, leading to rather narrow confidence intervals. This raises doubts about an infinite mean model and rather supports a model with a finite mean, but an infinite second moment.

\begin{table}
	\centering
	\begin{tabular}{cc|cc}
		threshold $u$ & $\#$ observations $\geq u$ & $\hat{t}_n(u)$ & corresponding $\alpha$ \\  \hline
		5  & 2197 & 0.44 & 0.82 \\
		25 &  571 & 0.43 & 0.85 \\
		60 &  293 & 0.40 & 0.94 \\
		150&  142 & 0.35 & 1.17 \\
		400&   50 & 0.34 & 1.18 \\
	\end{tabular}
	\caption{Values of $\hat{t}_n(u)$ for the wildfire suppression cost data for specific values of the threshold $u$ and the corresponding shape parameter $\alpha$ under the assumption that the tail data follow a Pareto model. }\label{tab3}
\end{table}

\begin{figure}
	\centering
	\includegraphics[width=\textwidth]{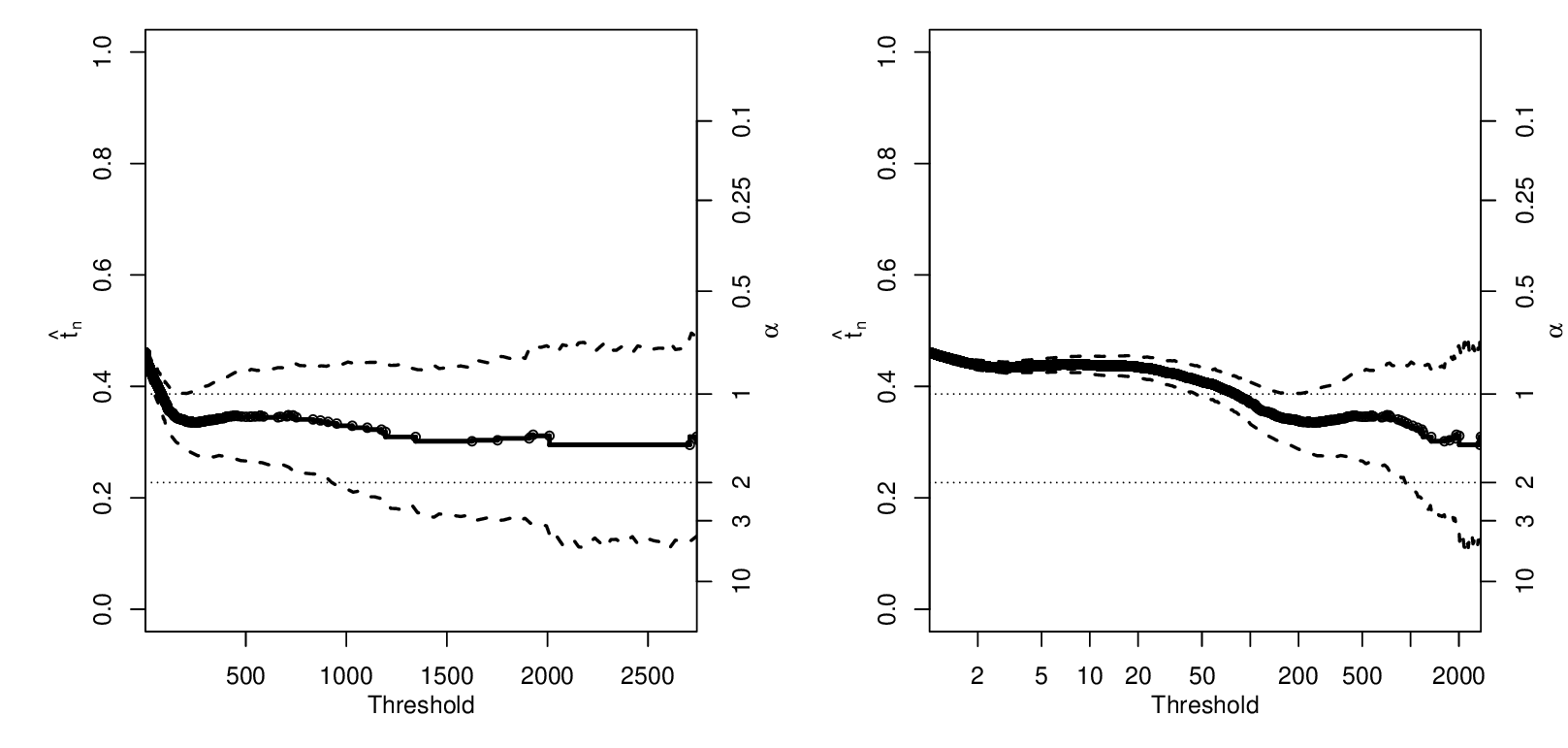}
	\caption{Plots for Wildfire suppression costs in subsection \ref{ss-wildfire}, $n=6599$. \\
		Left panel: Graph of $\hat{t}_n(u)$; the right scale indicates the corresponding shape parameter $\alpha$ under the assumption that the tail data follows a Pareto distribution.  \\
		Right panel: horizontal axis in log scale.
		\label{fig7}}
\end{figure}

\subsection{Costs of nuclear power accidents} \label{ss-nuclear}

\begin{figure}
	\centering
	\includegraphics[width=\textwidth]{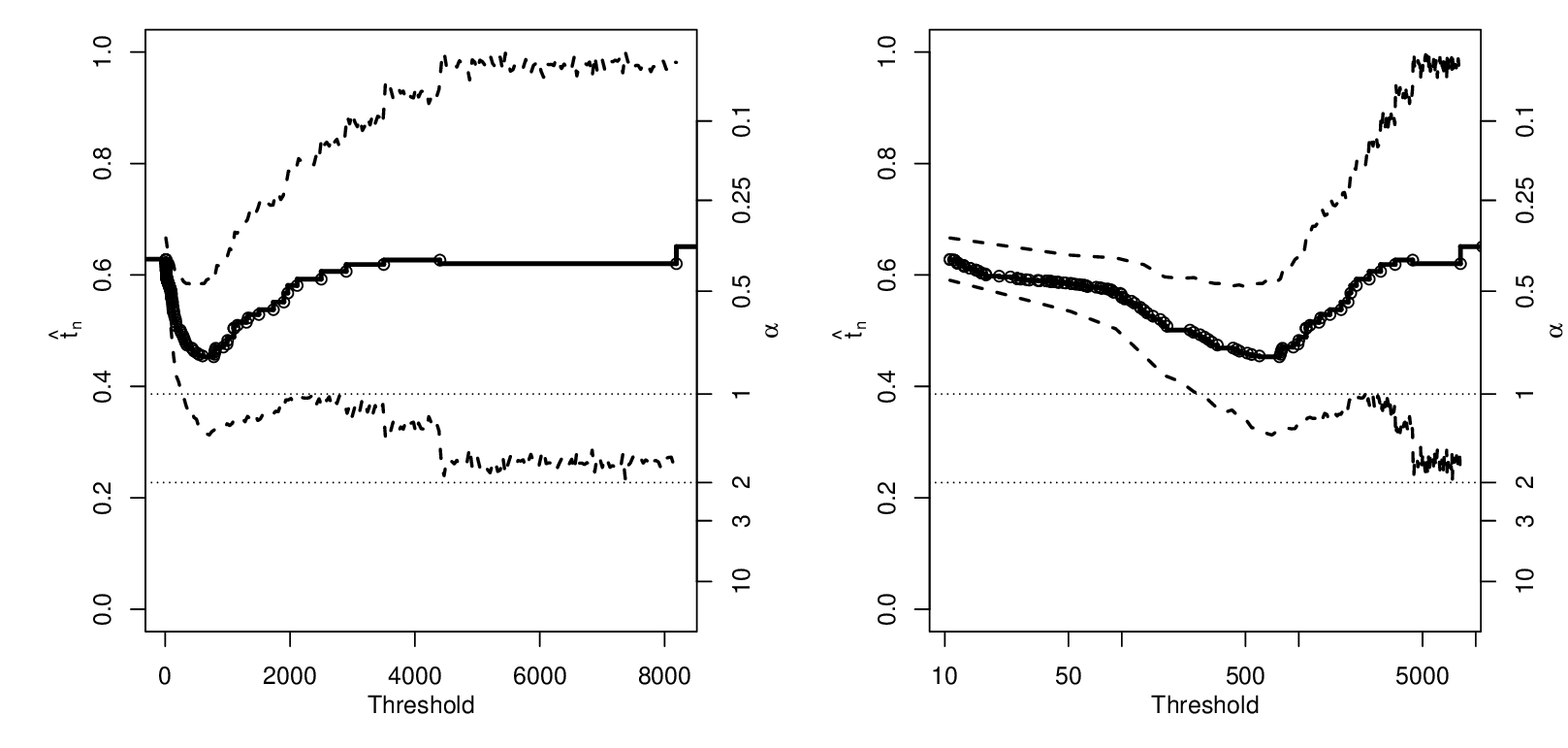}
	\caption{Plots for costs of nuclear power accidents in subsection \ref{ss-nuclear}, $n=125$. \\
		Left panel: Graph of $\hat{t}_n(u)$; the right scale indicates the corresponding shape parameter $\alpha$ under the assumption that the data follows a Pareto distribution.  \\
		Right panel: horizontal axis in log scale.
		\label{fig5}}
\end{figure}

A dataset of 216 incidents and accidents occurring at nuclear power plants, with costs in millions of U.S. dollars, is available under \url{https://data.world/rebeccaclay/nuclear-power-accidents}.
Considering only costs over 10 million dollars results in a dataset of size 125. The Pareto tail plot on the original scale is shown in the left panel of Figure \ref{fig5}; in the right panel, the horizontal axis is logarithmically scaled.
The Pareto tail function $\hat{t}_n$ takes values between $0.45$ and $0.63$, corresponding to values of $\alpha$ between $0.78$ and $0.40$. 
For this dataset, the graph does not really show stability, even in the tail. On the other hand, even when considering the full dataset, it is quite small, so that a Pareto tail is not necessarily excluded. 
Consider the specific thresholds $u=75,500,1640$, which roughly  correspond to the $0.5, 0.75$ and $0.90$ quantiles, using the $64, 31$ and $13$ largest data points. The pertaining values of $\hat{t}_n(u)$ are given by $0.58, 0.46,0.54$, corresponding to values of $\alpha$ given by $0.49, 0.76, 0.57,$ respectively. Taking into account the width of the confidence intervals in Figure \ref{fig5} shows that a Pareto distribution with tail function $t\equiv 0.5$, corresponding to a shape parameter around $0.65$, is a possible model in the tail. In any case, it is very likely that the underlying distribution has an infinite mean value.

\section{Concluding remarks}

This article proposes a graphical method for assessing the validity of a Pareto model, especially with respect to the tails of the distribution. It is a fully nonparametric approach, applicable to all continuous loss distributions without any moment restrictions. This is in sharp contrast to the mean excess plot, which plays a fundamental role in many fields, but is of limited value when dealing with heavy tails, although it is available in virtually all EVT software \citep[Table 4]{BD:2023}.
Furthermore, the proposed graphical approach can also be used as a {\em threshold stability plot}, where we choose the lowest threshold above which $\hat{t}_n$ is approximately constant, taking into account the uncertainty quantified by the interval estimates. In this respect, it would we worthwhile to construct simultaneous confidence bands; for the mean excess plot, such bands have been derived by \cite{DG:2013}.
The basic principle underlying the plot can also be interpreted as the {\em principle of a single huge jump}, distinguishing between distributions with moderate and super-heavy tails. Further research is needed to realize the full potential of this concept.


\section*{Acknowledgments}
I would like to thank Gerd Herzog for helpful discussions on this work. Furthermore, I would also like to thank two anonymous reviewers for their constructive and helpful comments.

\appendix

\section{Proof of Theorem \ref{pareto-char}} \label{appendix-A}

\begin{proof}
	For $n=2$, Srivastava's characterization yields that $X_{(1)}$ and $W=X_{(1)}/(X_{(1)}+X_{(2)})$ as well as $X_{(1)}$ and $1-W=X_{(2)}/(X_{(1)}+X_{(2)})$ are independent if and only if $X$ follows the Pareto distribution. Hence, the independence condition is equivalent to the condition that $X_{(1)}$ and $1-2W=(X_{(2)}-X_{(1)})/(X_{(1)}+X_{(2)})$ are independent, which yields the first assertion.
	
	Now, assume that $X,X_1,X_2\sim F(\cdot;\alpha)$ with density $f(\cdot;\alpha)$. The joint probability density function of $(X_{(1)},X_{(2)})$ is given by 
	\begin{align*}
		f(x_1,x_2) &= 2f(x_1;\alpha)f(x_2;\alpha)
		= 2\alpha^2 x_m^{2\alpha}
		\left(x_1 x_2\right)^{-(\alpha+1)}, \quad x_m\leq x_1 \leq x_2.
	\end{align*}
	It follows that $Y=(X_{(2)}-X_{(1)})/(X_{(1)}+X_{(2)})$ has density function
	\begin{align*}
		g(y) &= 2\alpha \, (1+y)^{-(\alpha+1)} (1-y)^{\alpha-1}, \quad 0\leq y<1.
	\end{align*}
	Thus, the computation of $\widetilde{t}_{\alpha}$ boils down to the evaluation of
	\begin{align}\label{h-par}
		\widetilde{t}_{\alpha} &= \Ex Y
		= \int_0^1 2\alpha \, (1+y)^{-\alpha} (1-y)^{\alpha-1}\, dy - 1
		= \check{t}_{\alpha}-1,
	\end{align}
	say. First, assume that $\alpha\in\mathbb{N}, \alpha\geq 2$. Using  integration by parts, we obtain
	\[
	\check{t}_{\alpha} = \frac{\alpha}{\alpha-1} \left( 2-\check{t}_{\alpha-1} \right),
	\]
	and a recursive application results in
	\[
	\check{t}_{\alpha} = 2\alpha \sum_{k=1}^{\alpha-1} \frac{(-1)^{\alpha+1-k}}{k} + (-1)^{\alpha+1} \,\alpha\, \check{t}_1.
	\]
	Since $\check{t}_1=2\log 2$, the first part of formula (\ref{t-pareto}) holds.
	Next, assume $\alpha\notin\mathbb{N}$. In this case, (\ref{t-pareto}) follows by noting that
	\begin{align*}
		\frac{\check{t}_{\alpha}}{2\alpha}
		&= \int_0^1 \, \frac{y^{\alpha-1}}{1+y}\, dy
		= \beta(\alpha),
	\end{align*}
	where $\beta(x)=\sum_{k=0}^{\infty} (-1)^k/(x+k) = (\Psi((x+1)/2)-\Psi(x/2))/2$ \citep[pp.\,292,\,947]{GR:2000}.
\end{proof}

\section{Proof of Theorem \ref{th-reg-var}} \label{appendix-B}

With $Z=(X_{(2)}-X_{(1)})/(X_{(1)}+X_{(2)})$, we obtain 
\begin{align*}
	t_X(u) &= \Ex\left[Z|X_{(1)} \geq u\right] 
	= \frac{\Ex\left[Z \cdot \ind\{X_{(1)} \geq u\}\right]}{P(X_{(1)} \geq u)}, 
\end{align*}
and, using the notation of Sec. \ref{sec-huge-jump},
\begin{align*}
	\Ex\left[Z \cdot \ind\{X_{(1)} \geq u\}\right] 
	&= \int_u^{\infty} \Ex\left[Z|X_{(1)}=s\right]  f_{X_{(1)}}(s) \, ds 
	= \int_u^{\infty} \int_0^1 z g(z|s) dz  \, f_{X_{(1)}}(s) \, ds.
\end{align*}
In the situation of Example \ref{ex-huge-jump}c), we have
$$
g(z|s)= \tilde{g}_{\alpha}(z) \, \frac{\tilde{L}\left(\frac{s(1+z)}{1-z} \right)}{L(s)}, \quad 0<z<1 \,.
$$ 
By assumption, for $\varepsilon>0$ there exists $s_0>x_m$ such that
\begin{align*}
	\left| \tilde{L}\big(s(1+z)/(1-z)\big) / L(s) - 1 \right|	&\, \leq \varepsilon \quad \forall s>s_0,
\end{align*}
independently of $z$. It follows that
\begin{align*}
	\left| \int_0^1 z g(z|s) dz - \int_0^1 z \tilde{g}_{\alpha}(z) dz \right|
	&\leq \varepsilon \int_0^1 z \tilde{g}_{\alpha}(z) dz \quad \forall s>s_0.
\end{align*}
Thus, for $u>s_0$,
\begin{align*}
	\left|t_X(u) - \tilde{t}_{\alpha}\right|
	&= \left| \frac{\int_u^{\infty} \int_0^1 z g(z|s) dz  \, f_{X_{(1)}}(s) \, ds}{P(X_{(1)} \geq u)} 
	- \int_0^1 z \tilde{g}_{\alpha}(z) dz \right| \\
	&= \left| \frac{\int_u^{\infty} \int_0^1 z \left( g(z|s)-\tilde{g}_{\alpha}(z) \right) dz  \, f_{X_{(1)}}(s) \, ds}{P(X_{(1)} \geq u)} \right| \\
	&\leq \varepsilon \int_0^1 z \tilde{g}_{\alpha}(z) dz,
\end{align*}
which proves the claim.


\section{Unbiased estimator for the variance of a U-statistic} \label{appendix2}

In this section, we discuss the estimation of $\Var(U_{n}^{(l)}(d)) (l=1,2)$ and $\Cov(U_{n}^{(1)}(d), U_{n}^{(2)}(d))$ 
by estimators $\sigma_{U^{(l)}}^2$ and $\sigma_{U^{(1,2)}}^2$ , respectively. 

Let $U_n=2/(n(n-1)) \sum_{i<j} h(X_i,X_j)$ be a general $U$-statistic of degree 2, estimating $\theta=\Ex h(X_1,X_2)$. Defining  $h_1(x_1)=\Ex h(x_1,X_2)$ and
\begin{align*}
\zeta_0=\theta^2, &\quad
\zeta_1=\Ex\left[h_1^2(X_1)\right],  \quad
\zeta_2=\Ex\left[h^2(X_1,X_2)\right],
\end{align*}
the finite sample variance of $U_n$ is given by
\[
	\Var(U_n) = \frac{2}{n(n-1)} \left\{ 2(n-2)\zeta_1 + \zeta_2 - (2n-3) \zeta_0 \right\}.
\]
One can estimate $\zeta_c, \, c=0,1,2,$ by
\begin{align*}
\hat{\zeta}_0 &= \frac{1}{n^{\underline{4}}} \sum_{d(i,j,k,l)} h(X_i,X_j)h(X_k,X_l), \nonumber \\
\hat{\zeta}_1 &= \frac{1}{n^{\underline{3}}} \sum_{d(i,j,k)} h(X_i,X_j)h(X_i,X_k), \qquad
\hat{\zeta}_2 = \binom{n}{2}^{-1} \sum_{i<j} h^2(X_i,X_j), 
\end{align*}
where $d(i_1,\ldots,i_m)$ denotes a set of distinct indices $1\leq i_1,\ldots,i_m \leq n$, and $n^{\underline{m}}=n(n-1)\cdots (n-m+1)$.
Then, the minimum variance unbiased estimator of $\Var(U_n)$ is given by \citep{SS:1992}
\begin{align} \label{unbiased-est}
\hat{\sigma}_U^2 &= \frac{2}{n(n-1)} \left\{ 2(n-2)\hat{\zeta}_1 + \hat{\zeta}_2 - (2n-3) \hat{\zeta}_0 \right\}
\ = \ U_n^2 - \hat{\zeta}_0.
\end{align}
The degree of the $U$-statistics in (\ref{unbiased-est}) is 4, but it is possible to rewrite it as
\begin{align} \label{unbiased-est2}
\hat{\sigma}_U^2 &= \frac{4C_1^2-2C_2^2}{n^{\underline{4}}}  - \frac{4n-6}{(n-2)(n-3)} U_n^2,
\end{align}
where
\begin{align*}
C_1^2 &= \sum_{i=1}^n \bigg( \sum_{j\neq i} h(X_i,X_j) \bigg)^2, \quad
C_2^2 = \sum_{i\neq j} h^2(X_i,X_j)
\end{align*}
\citep[p. 2972]{SS:1992}. 
To obtain a multivariate version of (\ref{unbiased-est2}), write $hh^T$ and $U_nU_n^T$ instead of $h^2$ and $U_n^2$, and define
\begin{align*} 
C_1^2 &= \sum_{i=1}^n S_iS_i^T, \quad \text{where }
S_i=\sum_{j\neq i} h(X_i,X_j).
\end{align*}

\bibliographystyle{apalike}
\bibliography{Pareto-tail-statistic-biblio}

\end{document}